\documentclass[12pt]{article}
\pdfoutput=1
\usepackage{jheppub}
\usepackage[utf8]{inputenc}
\usepackage{verbatim}
\usepackage{amsmath}
\usepackage{amssymb}
\usepackage{amsthm}
\usepackage{slashed}
\usepackage{amsfonts}
\usepackage{dsfont}
\usepackage[dvipsnames]{xcolor}
\usepackage{faktor}
\usepackage{tensor}
\usepackage{physics}
\usepackage{multirow}
\usepackage{stmaryrd} 

\newcommand{\D}{\textnormal{d}}
\newcommand{\be}{\begin{equation}}
\newcommand{\ee}{\end{equation}}
\newcommand{\bfig}{\begin{figure}\begin{center}}
\newcommand{\efig}{\end{center}\end{figure}}
\newcommand{\bi}{\begin{itemize}}
\newcommand{\ei}{\end{itemize}}

\newcommand{\A}{\mathcal{A}}

\theoremstyle{definition}

\newcommand{\M}{\mathcal{M}}
\newcommand{\R}{\mathbb{R}}
\newcommand{\N}{\mathbb{N}}
\newcommand{\C}{\mathbb{C}}
\newcommand{\Z}{\mathbb{Z}}
\newcommand{\e}{\varepsilon}

\newcommand{\rw}{\rightarrow}

\newcommand{\mc}{\mathcal}
\newcommand{\G}{\mc{G}}
\newcommand{\Gtilde}{\widetilde{\G}}
\newcommand{\Q}{\mathcal{Q}}
\newcommand{\im}[1]{\mathrm{im}\,#1}

\newcommand{\rrangle}{\mathclose{\rangle\mkern-4mu\rangle}}

\newcommand{\ri}{\rrbracket}

\newcommand{\rc}{\rrangle}
\newcommand{\lv}{\lvert}

\newcommand{\cket}[1]{\lv #1 \rc}

\newcommand{\iket}[1]{\lv #1 \ri}

\newcommand{\ham}{\mathcal{H}}
\newcommand{\Hkin}{\ham_{\text{kin}}}
\newcommand{\phys}{\text{phys}}
\newcommand{\Hphys}{\ham_\phys}

\newcommand{\Vco}{V_\text{co}}
\newcommand{\Vinv}{V_\text{inv}}
\newcommand{\Span}[1]{\text{Span}(#1)}

\newcommand{\kin}{\mathrm{kin}}

\newcommand{\vol}[1]{\mathrm{vol}(#1)}
\newcommand{\etamodk}{\tilde\eta_k}

\newcommand{\snpsi}[1]{\lVert #1\rVert_\psi}

\begin{document}
\title{The Hilbert space of gauge theories: group averaging and the quantization of Jackiw-Teitelboim gravity
}
\author[a]{Elba Alonso-Monsalve}
\affiliation[a]{Center for Theoretical Physics -- A Leinweber Institute\\ Massachusetts Institute of Technology, 77 Massachusetts Avenue, Cambridge, MA 02139, USA}
\emailAdd{elba\_am@mit.edu}
\abstract{When the gauge group of a theory has infinite volume, defining the inner product on physical states becomes subtle. This is the case for gravity, even in exactly solvable models such as minisuperspace or low-dimensional theories: the physical states do not inherit an inner product in a straightforward manner, and different quantization procedures yield a priori inequivalent prescriptions. This is one of the main challenges when constructing gravitational Hilbert spaces. In this paper we study a quantization procedure known as group averaging, which is a special case of the BRST/BV formalism and has gained popularity as a promising connection between Dirac quantization and gravitational path integrals. We identify a large class of theories for which group averaging is ill-defined due to isometry groups with infinite volume, which includes Jackiw-Teitelboim gravity. We propose a modification of group averaging to renormalize these infinite volumes and use it to quantize Jackiw-Teitelboim gravity with a positive cosmological constant in closed universes. The resulting Hilbert space naturally splits into infinite-dimensional superselection sectors and has a positive-definite inner product. This is the first complete Dirac quantization of this theory, as we are able to capture all the physical states for the first time.
}

\maketitle

\section{Introduction}\label{sec:intro}
Constructing the physical Hilbert space of a quantum theory with a gauge symmetry is an essential task in modern physics, yet depending on the nature of the gauge group it is sometimes unclear how to do this correctly. While the last decades have witnessed significant progress (see e.g.~\cite{henneaux1992quantization}), the discussion is still ongoing. One of the main challenges arises when the gauge group is non-compact, as this can cause some otherwise unproblematic quantization techniques to break down. A notable theory where this is an issue is gravity, since the group of diffeomorphisms on a smooth manifold is non-compact. Of course, gravity has additional problems, such as non-renormalizability or the impossibility of formulating it as a local quantum field theory without leading to the infamous loss of information. But even if one found a way to work around those other obstacles (for example, by working with low-dimensional toy models or minisuperspace reductions), the problem of a non-compact gauge group remains for universes like ours: in spacetimes without a spatial boundary (closed), time evolution is pure gauge.

Following the usual quantization procedure of Dirac \cite{diraclectures}, we often solve the equations of motion classically first---perhaps along with some of the constraints---and then apply the remaining constraints at the quantum level on the often-called ``kinematic'' or ``auxiliary'' Hilbert space $\Hkin$, in order to obtain the Hilbert space of physical states, $\Hphys$. We do this, for example, when we cannot solve all the classical constraints exactly. But this procedure is not without ambiguities, such as what we mean exactly by ``applying'' the constraints, how to choose factor ordering, or how to turn the resulting solutions into a Hilbert space. Different choices lead to a priori inequivalent quantization procedures, such as BRST/BV, path integrals, refined algebraic quantization (RAQ)\ldots~Ideally all these techniques should yield the same results, but comparing them is challenging, as they use different languages and do not always apply to the same theories.

An intuitive yet sometimes naive approach, which highlights one of the problems with a non-compact gauge group $G$, is the following: we let $\Hphys$ be the subspace of $\Hkin$ made up of states $\psi_\phys$ which are invariant under the action of $G$,
\begin{equation}\label{eigen1}
    U(g)\psi_\phys = \psi_\phys,
\end{equation}
for all $g\in G$, where $U(g)$ denotes the unitary representation by which $g$ acts on $\psi_\phys$. When $G$ is compact, this is unproblematic: $\Hphys$ contains all the physical states and even inherits the Hilbert space structure from $\Hkin$ (notably its inner product). However, when $G$ is non-compact, the spectrum of $U(g)$ might be too small. In particular, (\ref{eigen1}) might not have a solution (besides $\psi_\phys=0$), naively resulting in a 0-dimensional physical Hilbert space---too small to represent the physics which we actually measure. We will see examples in Sections \ref{sec:examples} and \ref{sec:JT}.

The problem this raises is twofold. Firstly---and more obviously---, this approach misses physical states. Secondly, if we cannot construct $\Hphys$ as a subspace of $\Hkin$, then it won't naturally inherit an inner product, and we must work harder to turn the space of physical states into a Hilbert space. Addressing the first problem will unavoidably land us in the second, so while the problem of the missing states gets alleviated by employing more sophisticated procedures, the problem of how to turn the space of physical states into a Hilbert space is quite pervasive. This lies at the core of existing ambiguities around how to define the inner product in quantum gravity.

One technique to tackle these problems which is growing in popularity is known as group averaging. This has been applied with reasonable success to many theories of interest, including in recent years \cite{Ashtekar:1995zh,   Balasubramanian:2025rcr, Held:2024rmg, Penington:2023dql, Marolf:2008hg}. Group averaging is a special case of RAQ, applicable when the gauge group is locally compact, Hausdorff and topological (so integration makes sense), such as a Lie group \cite{Giulini:1999kc}. The RAQ procedure, in very brief summary, involves constructing a map (called the rigging map, $\eta$) which directly maps (a dense subspace\footnote{We will discuss in detail the necessity of this dense subspace and how to construct it in Section \ref{sec:GAformalism}.} $\Phi$ of) $\Hkin$ into $\Hphys$ \cite{Marolf:1995cn,Giulini:1998rk}. The rigging map is then used to define an inner product on $\Hphys$. The main advantage of the group averaging version of RAQ is its practicality: the rigging map is defined as a reasonably uncomplicated integral over the gauge group (thus the name). This makes a lot of the physics manifest, making this quantization procedure both less obscure and easier to compute than others. Moreover, it has been shown \cite{Giulini:1998kf} that when this group-averaging integral is well defined,\footnote{Specifically, when it converges absolutely for a family of measures on the gauge group \cite{Giulini:1998kf} (which reduces to absolute convergence when the gauge group is unimodular).} the rigging map of RAQ is unique, and given explicitly by the group-averaging procedure. In the case of gravity, this is sometimes called the revised Wheeler-DeWitt formalism, and it can help formalize the cutting and gluing constructions (summing over intermediate states) employed in gravitational path integrals---although a rigorous connection is not yet fully understood (see \cite{Witten:2022xxp,Held:2025mai} for recent reviews). This quantization procedure can also be seen as a special case of BRST \cite{Shvedov:2001ai}.

These promising features motivate a closer study of group averaging. In this paper we examine this formalism for non-compact $G$. We identify a large class of theories for which the standard definition of the group averaging rigging map diverges, and thus is not well defined. The divergence we identify is caused by the volumes of non-compact subgroups of $G$ which act trivially on test functions (isometry groups, also known as stabilizer subgroups). We propose a prescription to ``renormalize'' the rigging map in these cases. Our renormalized group averaging prescription is quite general: it can be applied to any theory where $\Hkin = L^2(X)$ is the space of normalizable states on a measure space $X$ and $G$ is a finite-dimensional Lie group which acts on $X$ leaving its measure invariant and on $\Hkin$ via a unitary representation.\footnote{We will also assume $G$ is unimodular (there exists a Haar measure which is both left- and right-invariant). This is the case for all semisimple Lie groups. However, we believe our formalism should generalize to the case of non-unimodular $G$ (and quasi-invariant measure on $X$) in the manner of \cite{Giulini:1998kf}.} Our proposal also matches the standard group averaging when these divergences are not present.

Our proposed renormalization of the group-averaging integral also exposes an unavoidable obstacle to applying group averaging to some theories. Specifically, a direct corollary of our prescription is that, when $X$ has different orbit types associated to non-compact isometry groups (explained in Section \ref{sec:mod}), it is impossible to renormalize the aforementioned divergences on all of $\Phi\subset \Hkin$ at once. This signals that the physical Hilbert space of the corresponding quantum theory naturally splits into a sum over superselection sectors, each resulting from a different (renormalized) rigging map following our prescription. This phenomenon is not new: it has already appeared in specific examples, such as some theories with gauge groups $SO(n,1)$ and $SL(2,\R)$ \cite{Gomberoff:1998ms, Louko:1999tj, Louko:2005qj}. A split into sectors of a similar nature to what we present here was also observed in \cite{Marolf:1995cn,Ashtekar:1995zh} for abelian and compact gauge groups, respectively. However, to our knowledge, a general and predictive characterization of when and why this split into superselection sectors happens (for non-abelian, non-compact gauge groups), along with a prescription to construct $\Hphys$ in all cases, was missing. The inner product we obtain suffers from a (finite) normalization ambiguity across superselection sectors (the inner product can be rescaled independently on each sector). This ambiguity is universal, and is equivalent to the ambiguity in the trace normalization for Type II von Neumann algebras \cite{Sorce:2024pte}.

An important theory which suffers from this exact problem is Jackiw-Teitelboim (JT) gravity, a model in $1+1$ spacetime dimensions where the dynamical degrees of freedom are the metric and a scalar field known as the dilaton. Its low number of dimensions make calculations tractable, while still capturing many confusing aspects of quantum gravity, such as gravitational path integrals \cite{Moitra:2021uiv, Iliesiu:2020zld, Stanford:2020qhm, Ferrari:2024kpz}, proposed dualities \cite{Harlow:2018tqv,Saad:2019lba,Johnson:2019eik,Cotler:2019nbi}, and, in the case of a closed universe, the problem of time---including the issue of non-compact time evolution. In fact, JT gravity has helped identify or sharpen important puzzles pertaining to closed universes (e.g.~\cite{Usatyuk:2024mzs,Fumagalli:2024msi,Cotler:2025gui}). Building on our previous results \cite{Alonso-Monsalve:2024oii}, we find that our renormalized group averaging prescription is perfectly suited for the quantization of JT gravity in closed universes. Our results from group averaging are consistent with and more general than \cite{Held:2024rmg}, as we are able to quantize the entire theory, including all of its superselection sectors. In particular, for positive cosmological constant (dS-JT), the physical Hilbert space we obtain is
\begin{equation}
    \Hphys = \ham_{\substack{\text{expand/}\\ \text{crunch}}} \oplus \ham_\text{sing.} \oplus \left( \bigoplus_{n=1}^\infty \ham^n_\text{BH} \right),
\end{equation}
where $\ham_\text{expand/crunch}$ contains (superpositions) of states where the the dilaton expands or crunches uniformly throughout the spacetime, $\ham_\text{sing.}$ contains states with a conical singularity in the past or future, and $\ham^n_\text{BH}$ contains states with $n$ black holes.

This paper is structured as follows. In Section \ref{sec:GAformalism} we give a self-contained and pedagogical review of the group averaging formalism.\footnote{We avoid getting into details of the more general RAQ which are not necessary for our purposes. (See e.g.~\cite{Marolf:2000iq} for a complementary review.)} We illustrate this procedure by applying it to a quantum particle on a line. In Section \ref{sec:mod} we describe the aforementioned problem with group averaging and explain our proposal for renormalization. In Section \ref{sec:JT} we apply our proposal to JT gravity in closed universes with a positive cosmological constant and derive the physical Hilbert space. Finally, in Section \ref{sec:discussion} we comment on implications of our findings for gravity more generally and avenues for future work. Additionally, in Appendix \ref{app:posdef} we prove that the (renormalized) group-averaging inner product is positive-definite for the large class of theories studied in Section \ref{sec:mod}.

\section{Group averaging formalism}\label{sec:GAformalism}
In this section we give a pedagogical introduction to group averaging. First, we cover all the background necessary for the rest of the paper (the rigging map and the inner product). Then we review additional aspects of group averaging which are important but may be skipped on a first read (the test subspace and coinvariants). We finish by illustrating group averaging on a simple example: a free particle on a line with translations gauged.

Throughout this paper, $\Hkin$ will denote the kinematic Hilbert space (obtained from the quantization of the unconstrained theory), $\Hphys$ the Hilbert space of physical states, and the gauge group $G$ will be a locally compact Hausdorff topological group (so it has a Haar measure $\D g$)\footnote{We will also assume that $G$ is unimodular (so the Haar measure is both left- and right-invariant). There are subtleties when $G$ is not unimodular, but it is possible to take care of these in the context of group averaging \cite{Giulini:1998kf,Marolf:2000iq}.} which acts on $\Hkin$ by a continuous unitary representation.\footnote{In Section \ref{sec:mod} we will further restrict to the case where $G$ is a Lie group and $\Hkin=L^2(X)$ is the space of square-integrable (normalizable) states on a measure space $X$.}

\subsection{Rigging map, inner product, and invariants}
As described in Section \ref{sec:intro}, group averaging is a quantization procedure especially apt for theories 
where $\Hkin$ is too small to contain all the gauge-invariant states, which is often the case when the gauge group $G$ is non-compact.\footnote{Gauge-invariant states in $\Hkin$ must solve (\ref{eigen1}). A unitary representation of $G$ which does not vanish along the non-compact directions must be infinite dimensional (except for the trivial one). When $G$ is a connected, non-compact Lie group (often the case in physics), a nonzero solution to (\ref{eigen1}) cannot exist, since it would generate a nontrivial finite-dimensional unitary representation of $G$.} To construct $\Hphys$, we define the rigging map $\eta$. This is an antilinear map from a so-called ``test'' subspace $\Phi\subset\Hkin$ into its continuous dual $\Phi'$, which is the space of continuous linear functionals (distributions) on $\Phi$. The image of $\eta$ will then be interpreted as the space of physical states. The test subspace $\Phi$ must satisfy a few basic requirements: it must be dense in $\Hkin$ (to capture almost all of $\Hkin$), closed under the action of $G$ (that is, $U(g)\Phi=\Phi$ for all $g\in G$) and of observables of interest, and admit a topology such that $\eta$ is continuous\footnote{The author thanks Don Marolf and Daniel Harlow for pointing out the need to make $\eta$ continuous.} (we use this topology to define $\Phi'$). It is essential to work with a test subspace for the rigging map to be well defined and produce reasonable physical states. In particular, $\eta$ cannot be continuous on all of $\Hkin$, as its image would miss physical states when $G$ is non-compact, so it would not solve the problem we saw in Section \ref{sec:intro}. For now we will assume that $\Phi$ and its topology have been adequately chosen, and postpone a detailed discussion of this test subspace to Section \ref{sec:Phi}. The rigging\footnote{The name ``rigging'' is borrowed from the theory of rigged Hilbert spaces (see e.g.~\cite{delaMadrid:2005qdg} for a review). It has nothing to do with the meaning of ``rigging'' as dishonestly arranging a result---instead it is a nautical metaphor: the rigging of a ship is the structure of ropes that supports the masts and sails.} map is defined as
\begin{equation}\label{eta}
\begin{split}
    \eta : \Phi &\rightarrow \Phi'\\
    \psi &\mapsto \eta(\psi)
\end{split}
\end{equation}
with the distribution $\eta(\psi)$ defined by its action on test states $\chi\in \Phi$:
\begin{equation}\label{etachi}
    \eta(\psi)[\chi] \equiv \int_G\D g\, (\psi,U(g)\chi)_\kin.
\end{equation}
(Note that, as before, we choose to forego braket notation. We use $(-,-)_\kin$ to denote the inner product in $\Hkin$.) We have used square brackets to denote the argument of a distribution, as is customary. Thus, $\eta$ turns a kinematic test state $\psi$ into a distribution $\eta(\psi)$, which itself maps other test states $\chi\in\Phi$ to numbers: inner products with $\psi$ averaged over the action of the gauge group $G$.

The rigging map is well defined as long as the integral in (\ref{etachi}) converges absolutely. This is not always the case (see \cite{Marolf:2008hg} for an analysis). In Section \ref{sec:mod} we will see an important class of theories (including JT gravity) where group averaging fails due to a divergent integral, and we will propose a renormalization of $\eta$ to address this issue. Nonetheless, for the purposes of this section, we will assume that $\eta$ is well defined. We will work out an example where $G$ is non-compact and $\eta$ is well defined in Section \ref{sec:examples}.

Thanks to the invariance of the Haar measure $\D g$, it follows that $\eta$ is Hermitian on $\psi$ and $\chi$, in the sense that
\begin{equation}\label{hermit}
    \eta(\psi)[U(g)\chi] = \eta(U^\dagger(g)\psi)[\chi],
\end{equation}
for all $g\in G$, and also that $\eta$ is real, in the sense that $\eta(\psi)[\chi]=(\eta(\chi)[\psi])^\ast$. Moreover, it also follows from invariance of $\D g$ that
\begin{equation}\label{ginv}
    \eta(\psi)[U(g)\chi] = \eta(\psi)[\chi]
\end{equation}
for all $g\in G$, so the image of $\eta$ ($\im\eta$) consists of $G$-invariant continuous distributions. Not every invariant continuous distribution might be obtainable as a group average: for example, there might be singular invariant distributions (such as delta functions), which do not result from averaging test states in $\Phi$. We usually call invariant continuous distributions simply ``invariants,'' and denote the space of invariants by $\Vinv$:
\begin{equation}
    \Vinv := \{ \Psi\in\Phi'\:\vert\: \Psi[U(g)\chi]=\Psi[\chi] \text{ for all }g\in G, \chi\in\Phi \}.
\end{equation}Then,
\begin{equation}\label{Vinv}
    \im\eta \subset \Vinv,
\end{equation}
and this inclusion is usually strict. The fact that $\im\eta\neq\Vinv$ in general is a feature, not a bug. According to a uniqueness theorem by Giulini and Marolf, when $\eta$ is well-defined on a dense $\Phi$ (and given some mild assumptions), it is the unique rigging map of RAQ, and thus (\ref{ga-ip}) is the unique inner product in the quantum theory \cite{Giulini:1998kf}. In general, it is not possible to define an inner product on the larger space $\Vinv$.\footnote{$\Vinv$ is, in a sense, a ``rigged'' Hilbert space---an enlargement of $\Hphys$ which includes Dirac delta functions and other distributions which were not originally in $\Hphys$ (but is not a Hilbert space). See e.g.~\cite{delaMadrid:2005qdg} for a review of rigged Hilbert spaces.}

We use the rigging map $\eta$ to define the ``group-averaging'' inner product on $\im\eta$:
\begin{equation}\label{ga-ip}
    (\eta(\chi),\eta(\psi))_\phys := \eta(\psi)[\chi].
\end{equation}
Note that this inner product is linear in the second argument (and antilinear in the first), as is customary in physics:\footnote{This follows from the fact that $\eta:\Phi\rw \Phi'$ is antilinear (and thus $\eta(\psi):\Phi\rw\C$ is linear). If we had chosen $\eta$ to be linear instead, (\ref{ga-ip}) would actually remain unchanged.}
\begin{equation}
    (\alpha\eta(\chi),\beta\eta(\psi))_\phys = (\eta(\alpha^\ast\chi),\eta(\beta^\ast\psi))_\phys = \eta(\beta^\ast\psi)[\alpha^\ast\chi] = \alpha^\ast\beta\,\eta(\psi)[\chi].
\end{equation}
We have already shown that this inner product is Hermitian (\ref{hermit}). One must also prove that it is positive-definite. This has been shown to be true for many theories of interest \cite{Higuchi:1991tk, Higuchi:1991tm}. We will prove positive-definiteness for the class of theories we study in Section \ref{sec:mod}.

As long as the group-averaging inner product (\ref{ga-ip}) does not diverge and is positive-definite, we have successfully turned $\im\eta$ into a pre-Hilbert space (a vector space with an inner product). Upon taking its (Cauchy\footnote{Henceforth, every time we talk about completeness we will mean in the Cauchy sense: every Cauchy sequence has its limit point in the (completed) space. In fact, all the spaces we complete in this paper will be metric spaces, and all convergent sequences in a metric space are Cauchy.}) completion in the topology induced by the group-averaging inner product (\ref{ga-ip}), we obtain a Hilbert space:
\begin{equation}\label{Hphys}
    \Hphys := \overline{\im\eta},
\end{equation}
since the inner product is guaranteed to extend to the completion of a pre-Hilbert space by taking limits. This is the Hilbert space of physical states. Notice that $\overline{\im\eta}\subset \Vinv$ thanks to the continuity of $\eta$, so all states in $\Hphys$ are gauge-invariant.

In the remainder of this section, we will discuss additional background on group averaging which is not essential for the rest of the paper, so the reader may wish to skip to Section \ref{sec:examples} for an illustrative example, or straight to Section \ref{sec:mod}.

\subsection{Coinvariants}\label{sec:coinv}
In the previous subsection we constructed the physical states from the image of the rigging map $\eta$. Sometimes it is more convenient to construct physical states through a quotient. In particular, thanks to the first isomorphism theorem, we have
\begin{equation}\label{firstisoeta}
    \faktor{\Phi}{\ker\eta} \cong \im\eta.
\end{equation}
This quotient is the space of equivalence classes
\begin{equation}
    [\psi] = \{ \psi + \chi \:\vert\: \chi\in\ker\eta \},
\end{equation}
so it identifies states $\psi\in\Phi$ whose difference is a null state $\chi$ in the group-averaging inner product (\ref{ga-ip}). The vector-space isomorphism (\ref{firstisoeta}) is antilinear, and maps $[\psi]$ to $\eta(\psi)$. Thanks to (\ref{firstisoeta}), we can endow  equivalence classes in the quotient with the group-averaging inner product (\ref{ga-ip}), via
\begin{equation}\label{co-ip}
    ([\psi],[\chi])_\phys := \eta(\psi)[\chi],
\end{equation}
and then complete to obtain the physical Hilbert space $\Hphys$.\footnote{If we had defined $\eta$ to be linear instead of antilinear, we would have needed to flip $\psi$ and $\chi$ in (\ref{co-ip}).}

The quotient in (\ref{firstisoeta}) is different from another common quotient: the so-called space of ``coinvariants,'' $\Vco$. This is defined as
\begin{equation}\label{Vco}
    \Vco := \faktor{\Phi}{\overline{\Span{\mc{N}}}}
\end{equation}
where $\mc{N}$ is a subset of $\Phi$ defined as
\begin{equation}\label{null}
    \mc{N} := \{ U(g)\chi - \chi\:\vert\: g\in G, \chi\in\Phi \},
\end{equation}
$\Span{\mc{N}}$ is the smallest vector space containing $\mc{N}$ (that is, the space of all finite linear combinations of elements of $\mc{N}$), and the overline denotes the closure\footnote{Taken in the topology of $\Phi$, as discussed in Section \ref{sec:Phi}.} in $\Phi$. The quotient (\ref{Vco}) intuitively kills all states of the form $U(g)\chi - \chi$, by declaring an equivalence between all states whose difference is of the form $U(g)\chi - \chi$. Note that, in particular, it enforces an equivalence $U(g)\chi \sim \chi$ for all $g\in G$ and $\chi\in\Phi$, too. The elements of $\Vco$ are the equivalence classes,
\begin{equation}\label{coinvclass}
    [\psi] = \{ \psi + U(g)\chi - \chi \:\vert\: g\in G, \chi\in\Phi \}.
\end{equation}
There is an isomorphism \cite{Rudin1991}
\begin{equation}
    \Vco' \cong \Span{\mc{N}}^\circ,
\end{equation}
where $'$ denotes again the space of continuous linear functionals and $\Span{\mc{N}}^\circ$ is the continuous annihilator of $\Span{\mc{N}}$, that is, the space of continuous linear functionals which vanish on $\Span{\mc{N}}$:
\begin{equation}
    \Span{\mc{N}}^\circ := \{ \Psi \in \Phi' \:|\: \Psi[\chi]=0 \text{ for all } \chi\in\Span{\mc{N}} \}.
\end{equation}
Notice that every distribution $\Psi\in \Span{\mc{N}}^\circ$ satisfies
\begin{equation}
    \Psi\left[ \sum_n^N \left(U(g_n)\chi_n-\chi_n\right) \right] = 0
\end{equation}
for all $g_n\in G$, $\chi_n\in\Phi$ and $N\in\N$, and this is true if and only if
\begin{equation}\label{invdist}
    \Psi[U(g)\chi] = \Psi[\chi]
\end{equation}
for all $g\in G$ and $\chi\in\Phi$, so $\Span{\mc{N}}^\circ$ is the space of continuous $G$-invariant distributions (\ref{Vinv}). Therefore,
\begin{equation}
    \Vco' \cong \Vinv,
\end{equation}
so invariants are dual to coinvariants (thus the name\footnote{Although this nomenclature is unfortunately backward: according to the name, coinvariants should be dual to invariants (but for general infinite-dimensional vector spaces they are not).}). Invariants are often denoted by the symbol $\iket\psi$, and coinvariants are often denoted by $\cket\psi$ (for an arbitrary representative $\psi$ of the equivalence class (\ref{coinvclass})). It is common to say that the group-averaging inner product acts on invariants and coinvariants, but this is subtle: (\ref{ga-ip}) is only defined on $\Hphys\subset\Vinv$. Whenever we write the overlap of states in $\Vinv$ which are not in $\Hphys$, we mean it in the sense of rigged Hilbert spaces (see e.g.~\cite{delaMadrid:2005qdg} for a review).

Some readers might wonder why we did not define (\ref{Vco}) simply as the quotient of $\Phi$ by the equivalence relation induced by action of $G$, $U(g)\psi\sim \psi$. This is because the resulting quotient would not be a vector space. However, by quotienting by a subspace instead of by a group action, $\Vco$ becomes a vector space.

An alternative option might have been tempting: foregoing $\Phi$ and quotienting the full $\Hkin$ by the closure $\overline{\Span{\mc{N}}}$ taken in the topology induced by $(-,-)_\kin$. This is, at first glance, an appealing choice, because the quotient of a Hilbert space by a closed subspace is a Hilbert space too: it is Cauchy complete (Banach) and inherits a quotient norm which satisfies the parallelogram law, and thus induces an inner product. This would bypass ambiguities and evade the need of group averaging altogether. It sounds too good to be true and, in fact, it is, as we now show. By the Riesz representation theorem, there exists an isomorphism between a Hilbert space and its continuous dual:
\begin{equation}
    \faktor{\Hkin}{\overline{\Span{\mc{N}}}} \cong \left( \faktor{\Hkin}{\overline{\Span{\mc{N}}}}\right)'.
\end{equation}
As before, we also have an isomorphism
\begin{equation}
    \left( \faktor{\Hkin}{\overline{\Span{\mc{N}}}}\right)' \cong \Span{\mc{N}}^\circ_{\Hkin},
\end{equation}
where $\Span{\mc{N}}^\circ_{\Hkin}$ denotes the continuous annihilator of $\Span{\mc{N}}$ in $\Hkin'$:
\begin{equation}
    \Span{\mc{N}}^\circ_{\Hkin} := \{ \Psi \in \Hkin' \:|\: \Psi[\chi]=0 \text{ for all } \chi\in\Span{\mc{N}} \}.
\end{equation}
By analogous arguments to the ones used around (\ref{invdist}), this is the space of $G$-invariant continuous distributions in $\Hkin'$. But by the Riesz representation theorem, $\Hkin'\cong \Hkin$, and thus $\Span{\mc{N}}^\circ_{\Hkin}$ is equivalent to the space of $G$-invariant states in $\Hkin$. Therefore,
\begin{equation}\label{Hkinclosedspan}
    \faktor{\Hkin}{\overline{\Span{\mc{N}}}} \cong \{G\text{-invariant states in }\Hkin\},
\end{equation}
and, as we already discussed, this space is too small to capture the physical states when $G$ is non-compact.

\subsection{Test subspace $\Phi$}\label{sec:Phi}

We have insisted from the beginning on requiring that the rigging map $\eta$ be continuous The reason is that, when $\eta$ is continuous, it maps into $\Phi'$, as opposed to the larger algebraic dual of (not necessarily continuous) linear functionals.\footnote{If $\eta:\Phi\rw \im\eta$ is continuous with respect to the topology induced by (\ref{ga-ip}) on $\im\eta$, then $\eta(\psi):\Phi\rw\C$ is continuous too, and thus $\im\eta\subset\Phi'$. The converse is also true.} That space would be too large, as it contains distributions which are rather pathological, lacking desirable physical properties. For example, only continuous distributions have the property that if they vanish on a dense subspace, they vanish everywhere.

It is essential to define $\eta$ on a smaller test subspace $\Phi$ instead of all of $\Hkin$. If we constructed a rigging map which were continuous on all of $\Hkin$ (in the topology induced by the kinematic inner product), we would run into a similar problem as in the end of Section \ref{sec:coinv}: elements in the image of $\eta$ would be $G$-invariant elements in the continuous dual $\Hkin'$, but $\Hkin'$ is isomorphic to $\Hkin$,\footnote{By the Riesz representation theorem.} so $\im\eta$ would be a subspace of $\Hkin$ and thus too small to contain all the physical states when $G$ is non-compact. The key fact is the series of (strict) inclusions $\Phi\subset\Hkin\cong\Hkin'\subset \Phi'$.\footnote{In the language of rigged Hilbert spaces, $\{\Phi,\Hkin,\Phi'\}$ form a Gelfand triplet. The convention in the context of rigged Hilbert spaces is to work with the continuous antilinear dual $\Phi^\times$ as opposed to $\Phi'$, see e.g.~\cite{delaMadrid:2005qdg}. The preferred convention in group averaging differs because, by making $\eta$ antilinear (and thus a map into $\Phi'$ instead of $\Phi^\times$), the inner product (\ref{co-ip}) does not require us to flip $\psi$ and $\chi$.}

The basic requirements on $\Phi$ are that it be dense in $\Hkin$ (so it captures almost all of $\Hkin$), closed under the action of the gauge group ($U(g)\Phi=\Phi$ for all $g\in G$), and satisfy
\begin{equation}\label{L1G}
    (\psi,U(g)\chi)_\kin \in L^1(G),
\end{equation}
for all $\psi,\chi\in\Phi$. This ensures absolute convergence of the group-averaging integral (\ref{etachi}). When (\ref{L1G}) fails but we can renormalize the rigging map, we shall modify the third requirement accordingly, to ensure the renormalized rigging map converges. Sometimes it is also impossible to define $\Phi$ to be dense in all of $\Hkin$: when that happens, we must construct several test subspaces instead, each dense in a different disjoint subspace of $\Hkin$, and construct a rigging map as usual on each of the test subspaces. This naturally induces a split of $\Hphys$ into superselection sectors. This will be the case for the theories we study in Section \ref{sec:mod}.

The test subspace $\Phi$ inherits a topology from $\Hkin$ already (induced by the kinematic inner product), but if $\Phi\subset\Hkin$ is dense, then a continuous $\eta$ with respect to this ``kinematic'' topology on $\Phi$ would also be continuous on all of $\Hkin$, and thus map into $\Hkin'\cong\Hkin$ and suffer the problem we just described. We must put a finer topology on $\Phi$ to ensure continuity of the rigging map. We propose the following uncountable family of seminorms on $\Phi$, inspired by the requirement (\ref{L1G}):
\begin{equation}\label{sn}
    \snpsi{\chi} := \int_G\D g\,\left| (\psi,U(g)\chi)_\kin \right|,
\end{equation}
one for each $\psi\in\Phi$, where $|-|$ denotes the absolute value in $\C$. These induce a (locally convex) topology on $\Phi$: a sequence converges if it converges in all the seminorms. Let $\Delta:=\chi_0-\chi$ denote the difference between two test states in $\Phi$. If $\snpsi{\Delta}<\e$ for some $\e\in\R$, then by the triangle inequality for integrals,
\begin{equation}
    \left| \eta(\psi)[\Delta] \right| = \left| \int_G\D g (\psi,U(g)\chi)_\kin\right| \leq \snpsi{\Delta} < \e,
\end{equation}
and therefore $\eta(\psi):\Phi\rw\C$ is continuous, for all $\psi\in\Phi$. Thus, $\eta$ (\ref{eta}) is continuous too, with respect to the topology induced by the group-averaging inner product (\ref{ga-ip}) on its image.

This already specifies $\Phi$ and its topology, but there might be additional desirable requirements depending on what observables we want to be available in the quantum theory. We will borrow notions from the closely related theory of rigged Hilbert spaces \cite{delaMadrid:2005qdg}. Let $\A$ be an algebra of observables (self-adjoint operators on $\Hkin$ which commute with the action of $G$, and thus also with $\eta$) which we are interested in. Then we want $\Phi$ to be a subspace on which the observables yield meaningful values. In particular, we want the test subspace to be closed under the action of operators in $\A$. To this end, we can define a smaller test subspace
\begin{equation}\label{Phi}
    \Phi_\A := \{ \psi\in\Phi\:\big\vert\: A\psi\in \Phi \text{ for all }A\in\A \}.
\end{equation}
Assume $\A$ is generated by a countable subalgebra $\{A_1,A_2,\cdots\}$. There is a natural family of seminorms on $\Phi_\A$:
\begin{equation}\label{snA}
    \lvert \psi \rvert_{\{n_i\}} := |A_1^{n_1}A_2^{n_2}\cdots\psi|_\kin,
\end{equation}
where $|\cdots|_\kin$ is the norm induced by the kinematic inner product. Together with (\ref{sn}), these induce a finer topology on the (smaller) test subspace.

\subsection{Example: particle on a line}\label{sec:examples}
Now we apply group averaging to a simple theory: a particle on a line, with translations gauged. This will serve as a more concrete introduction to this quantization procedure. The kinematic Hilbert space in position-space polarization is the space of normalizable (square-integrable) wavefunctions $f:\R\rw\C$,
\begin{equation}
    \Hkin := L^2(\R),
\end{equation}
and the gauge group $G\cong \R$ is the group of translations. The space $\Hkin$ is Hilbert with respect to the $L^2$ inner product,
\begin{equation}
    (f,h)_\kin = \int_\R\D x \,f^\ast(x)\,h(x),
\end{equation}
where $^\ast$ denotes complex conjugation. We will henceforth drop the label $\R$ when the domain of integration is $\R$. Gauge transformations act as
\begin{equation}
    U(g)f(x) = f(x-a)
\end{equation}
for $g\in G$ which translates $f$ by an amount $a$. The naive space of physical states is the subspace of translation-invariant states---in other words, constant functions---in $\Hkin$. But nonzero constants are not normalizable, so the (naive) physical subspace is
\begin{equation}
    \{G\text{-invariant states}\} = \{0\},
\end{equation}
which is too small. We turn to group averaging to solve this problem.

Let $\Phi\subset\Hkin$ be a suitably chosen test subspace, for example, the space of Schwartz functions $\mc{S}(\R)$.\footnote{$\mc{S}(\R)$ is the test subspace for the algebra generated by the position $X$, momentum $P$, and Hamiltonian $H$ operators for a quantum-mechanical particle \cite{SteinShakarchi2003,Schwartz1950}.} The rigging map is
\begin{equation}\label{etaR}
\eta(f)[h] = \int_G\D g\, (f,U(g)h)_\kin = \int\D a\int\D x\,f^\ast(x)\,h(x-a).
\end{equation}
If $\Phi\subset L^1(\R)\cap L^2(\R)$ or, in other words, the test functions are chosen to be Lebesgue integrable, we can commute the integrals.\footnote{Proof: let $f,h$ be Lebesgue integrable. Consider the integral $\int\D a\int\D x\,\lvert f^\ast(x)\,h(x-a)\rvert.$ The integrand is everywhere nonnegative, so by Tonelli's theorem we can commute the integrals. Using the fact that the measure is invariant under translations and changes of sign (inversions), $\int\D a\int\D x\,\lvert f^\ast(x)\,h(x-a)\rvert=\int\D x\int\D a\,\lvert f^\ast(x)\,h(x-a)\rvert = \int\D x\,\lvert f(x)\rvert\int\D a\,\lvert h(x-a)\rvert=\left(\int\D x\,\lvert f(x)\rvert\right)\left(\int\D a\,\lvert h(a)\rvert\right)$. This is the product of two integrals which are finite by assumption, so we conclude that $\eta(f)[h]$ converges absolutely. Thus, by Fubini's theorem, we may exchange the integrals.} This is the case for all functions in $\mc{S}(\R)$.\footnote{So $\mc{S}(\R)$ satisfies (\ref{L1G}) too.} Then, using invariance of the measure under translations and inversions, we obtain
\begin{equation}\label{etaR}
\eta(f)[h] = \left(\int\D x\,f^\ast(x)\right)\left(\int\D x\,h(x)\right).
\end{equation}
Each of the integrals yields a finite number,
\begin{equation}\label{cf}
    c_f^\ast := \int\D x\,f^\ast(x), \quad c_h := \int\D x\,h(x),
\end{equation}
with $c_f,c_h\in\C$, so we see that $\eta(f)$ converges to a ``constant'' distribution (a distribution with a constant kernel $c_f^\ast$):
\begin{equation}
    \eta(f)[h] = \int\D x\,c_f^\ast h(x) = c_f^\ast c_h.
\end{equation}
Thus, $\im\eta\subset\Vinv$. We can easily pick $f\in\Phi$ such that it integrates to any choice of $c_f^\ast\in\C$, so we conclude that, in this theory specifically, $\im\eta=\Vinv$. Notice that this result is independent of the choice of test subspace $\Phi$, so long as we choose it to be Lebesgue. The group-averaging inner product (\ref{ga-ip}) is therefore the scalar product of complex numbers,
\begin{equation}\label{ga-ip-R}
    (\eta(f),\eta(h))_\phys = c_f^\ast c_h.
\end{equation}
Upon completing $\im\eta$, we obtain the Hilbert space of physical states $\Hphys$ (\ref{Hphys}). The inner product (\ref{ga-ip-R}) is the simplest inner product which turns $\Hphys\cong\C$ into a Hilbert space.

\section{A renormalized group averaging proposal}
\label{sec:mod}

Now that we have reviewed group averaging and appreciated its benefits, we turn to an large class of theories that probe its limitations. In this section, we show that whenever a non-compact subgroup of $G$ acts trivially on enough states in $\Hkin$, the standard rigging map (\ref{eta}) is doomed to diverge, and therefore we cannot use it to define the physical Hilbert space. This is generally the case in theories of gravity, where some metric configurations have isometries.\footnote{Field theories lie beyond the scope of this paper, but generalizations of group averaging to infinite-dimensional gauge groups may be possible via path integrals. We comment on this in Section \ref{sec:discussion}.} We then propose a scheme to renormalize the rigging map that works for all such theories. In Section \ref{sec:JT}, we will apply this to JT gravity, a notable example which suffers from this problem.

\subsection{Problem: non-compact stabilizers}

Consider a kinematic state $\psi$ such that $U(g)\psi = \psi$
for all $g$ in a subgroup of $G_\psi\subset G$ (note this is different from (\ref{eigen1})). That is, $G_\psi$ is the subgroup of isometries of $\psi$. If $G_\psi$ is non-compact, then the absolute value of the group-averaging integral is bounded below for all $\chi\in\Phi$, by
\begin{equation}
    \int_{G}\D g\,\left|(\psi,U(g)\chi)_\kin\right| \geq \int_{G_\psi}\D g\,\left|(\psi,U(g)\chi)_\kin\right| = \vol{G_\psi}\,(\psi,\chi)_\kin,
\end{equation}
where
\begin{equation}\label{volstab}
    \vol{G_\psi}=\int_{G_\psi}\D g \rw \infty
\end{equation}
is the volume of the non-compact $G_\psi$, which diverges. Thus, the rigging map (\ref{etachi}) does not converge absolutely (in other words, the requirement (\ref{L1G}) fails). If this is true for too many $\psi$, in the sense that it is impossible to choose a dense $\Phi\subset\Hkin$ which does not contain one such $\psi$, then $\eta$ cannot be well defined, and we must modify the group averaging procedure so as to renormalize the divergence from (\ref{volstab}).

In order to make calculations concrete, in the rest of this paper we will restrict to the common case where $G$ is a finite-dimensional\footnote{So it is locally compact and thus has a Haar measure.} Lie group and $\Hkin=L^2(X)$ consists of normalizable functions on a (locally compact, Hausdorff, and second-countable) topological space $X$. A usual example is any theory where the classical pre-phase space (the space of classical solutions before applying the constraints, also known as unreduced phase space) is a manifold with a cotangent bundle structure $T^\ast X$ (then, in the standard position-space polarization, $\Hkin=L^2(X)$).\footnote{Pathologies which break all or part of this nice structure often only arise on the (reduced) phase space, that is, after one has applied the constraints at the classical level. A recent example where the phase space is neither a manifold nor Hausdorff was found in \cite{Alonso-Monsalve:2024oii}.} We will take $X$ to have a measure (e.g.~a Radon measure\footnote{This is a generalization of Haar (and Lebesgue) measures.}) $\D x$ invariant under $G$, with
\begin{equation}\label{unitarity}
    U(g^{-1})\psi(x) = \psi(g\cdot x),
\end{equation}
where $g\cdot x$ denotes the action of $G$ on $x\in X$ (which need not be left multiplication). This ensures unitarity on $\Hkin$, in the sense that $|U(g)\psi|_\kin = |\psi|_\kin$, where $|\ldots|_\kin$ denotes the $L^2$ norm. These assumptions are true for the particle on a line that we saw in Section \ref{sec:examples}, the theories studied in \cite{Gomberoff:1998ms,Louko:1999tj,Louko:2005qj}, and JT gravity, among many other theories.\footnote{We expect that our results in the next section should generalize to non-unimodular $G$ and quasi-invariant $\D x$ along the lines of \cite{Giulini:1998kf}, by relating the Radon-Nikodym derivative to the modular functions on stabilizer subgroups.} We will also require test subspaces $\Phi$ to be such that the rigging map converges absolutely as an integral over $G$ and $X$, which will allow us to exchange the order of integration, as we did in the example in Section \ref{sec:examples}.

The stabilizer $G_x$ of a point $x\in X$ (also sometimes called its isometry group) is defined as the subgroup of $G$ which leaves $x$ unchanged:
\begin{equation}
    G_x:= \{g\in G\:\vert\: g\cdot x=x\},
\end{equation}
When $G$ is non-compact, there might be some $x\in X$ whose stabilizer $G_x$ is also non-compact. Let $Y\subset X$ be the subset of such $x$. Now consider the rigging map on $\psi,\chi\in\Phi$ with support on $Y$. Then,
\begin{equation}\label{eta1}
    \eta(\psi)[\chi] = \int_X\D x\,\psi^\ast(x)\int_G\D g\,\chi(g\cdot x).
\end{equation}
For concreteness, choose $\psi,\chi$ which are everywhere positive. Now we see the issue: if $Y$ has positive measure in $X$, then the integration over $G$ yields an infinite volume. Specifically, (\ref{eta1}) is bounded below by the integral over $Y$ only,
\begin{equation}\label{intstab}
    \int_Y\D x\,\psi^\ast(x)\int_G\D g\,\chi(g\cdot x)=\int_Y\D x\,\psi^\ast(x)\, \vol{G_x}\int_{G/G_x}\D \dot{g}\,\chi(g\cdot x),
\end{equation}
where $\D\dot{g}$ is the quotient measure on $G/G_x$ ($\dot{g}$ denotes a representative of an equivalence class $[g]$). This quotient is the space of equivalence classes
\begin{equation}
    [g] = \{ gh\in G\:\vert\: h\in G_x \}.
\end{equation}
Note that $\vol{G_x}$ is infinite for all $x\in Y$, so the rigging map is doomed to diverge, and the divergence is parameterized by the volumes of the non-compact stabilizers. Note that in (\ref{intstab}) we have split the integral over the group $G$ into an integral over the stabilizer subgroup $G_x$ and its complement $G/G_x$. This split is guaranteed to exist by our assumptions on $G$ and $X$. Moreover, $\D \dot{g}$ is Radon, $G$-invariant, and unique (up to a constant rescaling).\footnote{See e.g.~Theorem 1.5.3 in \cite{DeitmarEchterhoffHA}.}

\subsection{Solution: renormalized rigging map}

We have conveniently parameterized the divergence from $\vol{G_x}$ of the rigging map as in (\ref{intstab}). To get rid of this divergence, we propose a ``renormalized'' group-averaging rigging map where we do not integrate over $G_x$ in the first place. This requires some care. The points $x\in Y$ need not all have isomorphic stabilizers.\footnote{For example, the stabilizers cannot be isomorphic if their orbits under the action of $G$ are not themselves isomorphic, due to the orbit-stabilizer theorem.} If $Y$ is measure zero in $X$, we can proceed with traditional group averaging.\footnote{This is because $L^2$ functions are defined up to an equivalence relation: two functions are equal if they coincide almost everywhere in $X$, so their values on a measure-zero subset of $X$ do not matter.} Otherwise, we split $Y$ into (disjoint) positive-measure\footnote{Measurability is guaranteed for second-countable $G$ and $X$, such as all finite-dimensional Lie groups and most smooth manifolds.} sets $Y_i$,
\begin{equation}\label{sumY}
    Y = \bigsqcup_i Y_i\quad\text{a.e.}
\end{equation}
such that all $x,y\in Y_i$ have stabilizers in the same conjugacy class, $G_x=gG_yg^{-1}$ for some $g\in G$. This conjugagy class of stabilizers is known as the ``orbit type,'' and $Y_i$ is often called the ``orbit type stratum''; all points in $Y_i$ have the same orbit type. The equality in (\ref{sumY}) holds almost everywhere (``a.e.''), that is, up to a set of measure zero. Let the index\footnote{There are at most countably many orbit type strata $Y_i$ with positive measure in $X$, thanks to the fact that $G$ preserves a $\sigma$-finite measure on $X$ \cite{Folland2015}. This is not true for measure-zero orbit type strata in $X$.} $i$ in (\ref{sumY}) start at $1$, and let $Y_0\in X$ denote the set of all $x$ with compact stabilizers. Almost all\footnote{A dense subset.} points in $Y_0$ actually have the same orbit type (the ``principal'' orbit type), so this notation is unambiguous.\footnote{This is thanks to the principal orbit type theorem \cite{Bredon1972}. This theorem applies to compact orbits even when $G$ is non-compact by restricting to the action of the maximal compact subgroup of $G$ (which contains all compact stabilizers) on $Y_0$.} Clearly $X=Y_0\sqcup Y$ a.e. Let the index $k$ start at 0, so
\begin{equation}\label{Xpart}
    X=\bigsqcup_kY_k\quad\text{a.e.}
\end{equation}
Since the $Y_k$ are disjoint by construction---and thus orthogonal in the $L^2$ inner product---(\ref{Xpart}) gives a measure space partition of $X$, so $\Hkin=L^2(X)$ decomposes into a direct sum
\begin{equation}\label{Hkinsplit}
    \Hkin = \bigoplus_k L^2(Y_k)
\end{equation}
with $L^2(Y_k)$ defined with respect to the measure on $X$ restricted to each $Y_k$. In order to apply group averaging, we restrict to test functions which have support only on one of the $Y_k$, that is, functions in $L^2(Y_k)$. Let $\Phi_k\subset L^2(Y_k)$ be suitable test subspaces, chosen as described in Section \ref{sec:examples}, with respect to the same algebra of observables if needed. With the test subspaces $\Phi_k$ chosen in this physically meaningful manner (due to a natural split of $\Hkin$), we expect group averaging to yield a reasonable $\Hphys$ and a sufficiently unique rigging map (after our renormalization below), in the sense of \cite{Marolf:1995cn}.

Thanks to the $G$-invariance of $\D x$, the stabilizers $G_x$ have equal volumes for all $x\in Y_k$. We then propose the following renormalized rigging maps $\etamodk$:
\begin{equation}
\begin{split}
    \etamodk : \Phi_k &\rightarrow \Phi_k^\ast\\
    \psi &\mapsto \etamodk(\psi),
\end{split}
\end{equation}
defined by
\begin{equation}\label{etamod}
    \etamodk(\psi)[\chi] := \int_{Y_k}
\D x\,\psi^\ast(x) \int_{G/G_x}\D \dot{g}\,\chi(g\cdot x),
\end{equation}
where $\D\dot{g}$ is again the $G$-invariant measure on $G/G_x$, unique up to rescaling.
Then the physical Hilbert space is a direct sum over superselection sectors
\begin{equation}\label{Hphysmod}
    \Hphys := \bigoplus_k \Hphys^k,
\end{equation}
where the Hilbert space in each sector is
\begin{equation}\label{Hphysk}
    \Hphys^k := \overline{\im{\etamodk}},
\end{equation}
with the closure of $\im{\etamodk}$ taken in the topology induced by the (renormalized) group-averaging inner product:
\begin{equation}\label{modip}
    (\etamodk(\chi),\etamodk(\psi))_\phys := \etamodk(\psi)[\chi].
\end{equation}
This inner product makes $\Hphys^k$ Hilbert spaces. By $G$-invariance of $\D\dot{g}$ and $\D x$, (\ref{modip}) is Hermitian, real, and the image of $\etamodk$ consists of $G$-invariant continuous distributions on $\Phi_k$. Moreover it is also positive-definite; we prove this in Appendix \ref{app:posdef}. We also give there a simpler form of $\etamodk$ (\ref{etamodsimp}).

By following the rules of Section \ref{sec:Phi} to construct $\Phi_k$, the rigging maps $\etamodk$ commute with the observables, and the corresponding quantum operators
\begin{equation}
    A_\phys \etamodk(\psi) := \etamodk(A\psi)
\end{equation}
map $\Hkin^k$ into $\Hkin^k$, and thus provide a superselection rule. As usual in theories with superselection sectors, there is a universal ambiguity in the normalization of the inner product across sectors: we can rescale $\etamodk$ (equivalently $\D \dot{g}$) by a different finite value on each sector $\Hphys^k$. This is the same as the ambiguity in the trace normalization for Type II von Neumann algebras (see \cite{Sorce:2024pte} for a recent review). Through our renormalized group averaging, we have traded a type III von Neumann algebra of gauge transformations on all of $L^2(X)$ for type II algebras on each $L^2(Y_k)$.

Let us now directly compare the standard and the renormalized rigging maps acting on test functions $\psi,\chi\in \Phi_k$, which have support only on $Y_k$. Since the volumes of all stabilizers of points in $Y_k$ are equal,
\begin{equation}\label{etavoletamod}
    \eta(\psi)[\chi] = \vol{G_x}\,\etamodk(\psi)[\chi],
\end{equation}
so we have successfully factored out the divergence we found in (\ref{intstab}), parameterized by the volume of the non-compact stabilizers.

By (\ref{etavoletamod}), we see that when $G$ is compact our ``renormalized'' prescription gives the same $\Hphys$ and inner product (up to rescaling with no physical implications) as standard group averaging. For the particle on a line from Section \ref{sec:examples}, the only translation that keeps points in $\R$ unchanged is a translation by 0 (the group identity $0\in G$), so all points $x\in\R$ have equal stabilizers, $G_x=\{0\}$ for all $x\in\R$, and there is a single orbit type. Thus, our renormalized group averaging gives the exact same $\Hphys$ as the old group averaging in this case too.

\section{Renormalized group averaging for JT gravity in closed universes}\label{sec:JT}

In this section, we use our renormalized group averaging from Section \ref{sec:mod} to complete the quantization of Jackiw-Teitelboim (JT) gravity with positive cosmological constant (dS-JT) in closed universes, which was kick-started by \cite{Alonso-Monsalve:2024oii}. This method is the only quantization procedure to date which can capture the entire Hilbert space (all the superselection sectors under the same framework). We start by reviewing\footnote{We encourage the reader to consult \cite{Alonso-Monsalve:2024oii} for more detailed explanations.} the phase space of classical solutions of dS-JT gravity on Cauchy slices with circle topology, which was first found by \cite{Alonso-Monsalve:2024oii} and satisfies the assumptions outlined above (\ref{unitarity}). Then we describe the orbit type strata $Y_k$ and construct the renormalized rigging map (\ref{etamod}). Finally, we write the physical Hilbert space, which splits into the superselection sectors described in the introduction.

\subsection{Unreduced phase space}
\begin{figure}
\centering
    \includegraphics[width=0.9\linewidth]{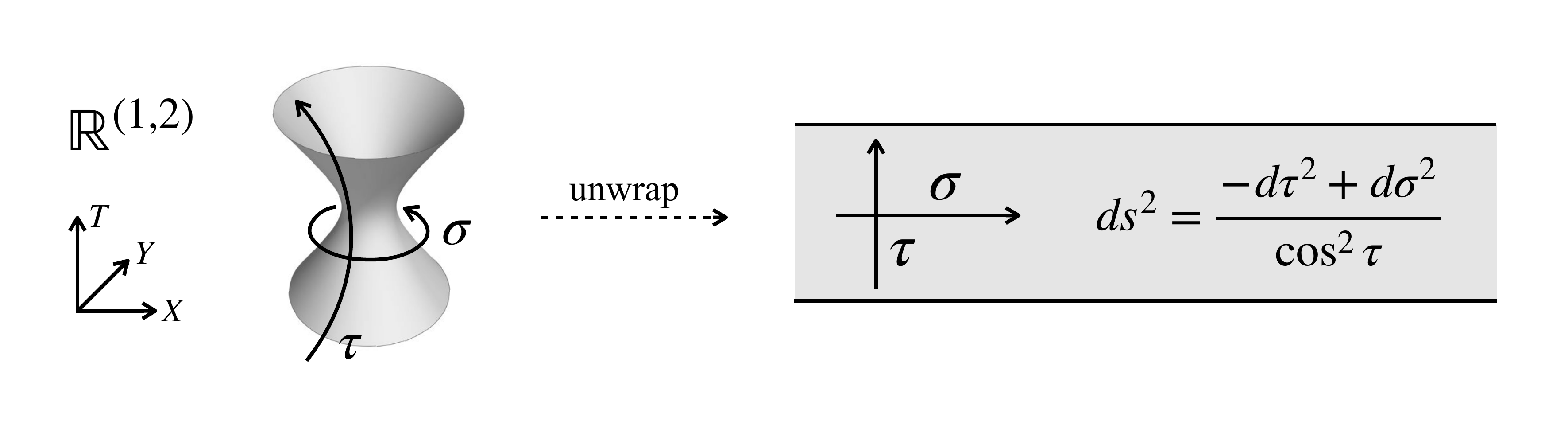}
\caption{The infinite strip, with $\tau\in(-\pi/2,\pi/2)$ and $\sigma\in\R$, is the universal cover of the de Sitter hyperboloid, (\ref{hyperboloid}). Figure borrowed from \cite{Alonso-Monsalve:2024oii}.}
\label{fig:hyp}
    \label{fig:placeholder}
\end{figure}

The action for JT gravity in spacetimes without a spatial boundary is (see e.g.~\cite{Mertens:2022irh} for a review)
\begin{equation}
S = \phi_0\int_\M\D x^2\sqrt{-g}R + \int_\M\D x^2\sqrt{-g}\,\Phi(R-2),
\label{JTaction-no-bdry}
\end{equation}
where $R$ is the Ricci scalar for the spacetime metric $g_{\mu\nu}$ and $\Phi$ is a scalar field called the dilaton. These are the dynamical degrees of freedom. The quantity $\phi_0$ is a large positive constant. We have chosen units such that the de Sitter length is 1, and focus on the pure theory with no additional matter. The equations of motion are
\begin{equation}\label{eoms}
\begin{split}
    R&=2,\\
    (\nabla_\mu\nabla_\nu+g_{\mu\nu})\Phi &= 0.
\end{split}
\end{equation}
The first equation constrains the metric locally to take the form
\begin{equation}\label{dsglobal}
    \D s^2 = \frac{-\D\tau^2 + \D\sigma^2}{\cos^2\tau}
\end{equation}
up to diffeomorphism, so solutions can only differ in their global structure. We can therefore construct solutions by starting with the universal cover of (\ref{dsglobal}) and taking quotients which yield Cauchy slices with circle topology. In order for the metric to wrap around the spacetime smoothly, we must quotient only by isometries $q$ (that is, we identify $x\sim q\cdot x$ for all points $x$ on the infinite strip). The universal cover has $\tau\in(-\pi/2,\pi/2)$ and $\sigma\in\R$, so we will henceforth refer to it as the ``infinite strip.'' The isometry group of the infinite strip is
\begin{equation}
    \Gtilde = \widetilde{PSL}(2,\R).
\end{equation}
This is the universal cover of the identity component of the 3-D Lorentz group,
\begin{equation}
    \G := PSL(2,\R) \cong SO^+(2,1).
\end{equation}
Equivalently, $\Gtilde$ is a central extension of $\G$ resulting from including translations by $2\pi n$ along the $\sigma$-direction, for all $n\in\N$. Intuition behind this isometry group follows from the familiar de Sitter hyperboloid
\begin{equation}\label{hyperboloid}
    -T^2 + X^2 + Y^2 = 1
\end{equation}
embedded in 3-D Minkowski space, $\R^{(1,2)}$, with metric
\begin{equation}
    \D s^2 = -\D T^2 + \D X^2 + \D Y^2.
\end{equation}
Note that, by setting
\begin{equation}
    T = \tan\tau,\, X = \frac{\cos\sigma}{\cos\tau},\, Y = \frac{\sin\sigma}{\cos\tau},
\end{equation}
we recover (\ref{dsglobal}) as the induced metric on the hyperboloid (\ref{hyperboloid}). Translations along the $\sigma$ direction correspond to rotations around the $T$-axis (see Figure \ref{fig:hyp}); in particular, a $2\pi$-translation in the $\sigma$-direction is equal to the identity. The hyperboloid (\ref{hyperboloid}) inherits all the isometries from the ambient Minkowski space, so its isometry group is the full Lorentz group. We will not be interested in quotienting by time reversal or parity transformations, to keep spacetime orientable, so we will focus only on the component connected to the identity, $\G$. The infinite strip is the universal cover of the hyperboloid: it wraps infinitely around the $\sigma$-direction, so its isometry group is $\Gtilde$.

\begin{figure}
\centering
\includegraphics[width=0.45\linewidth]{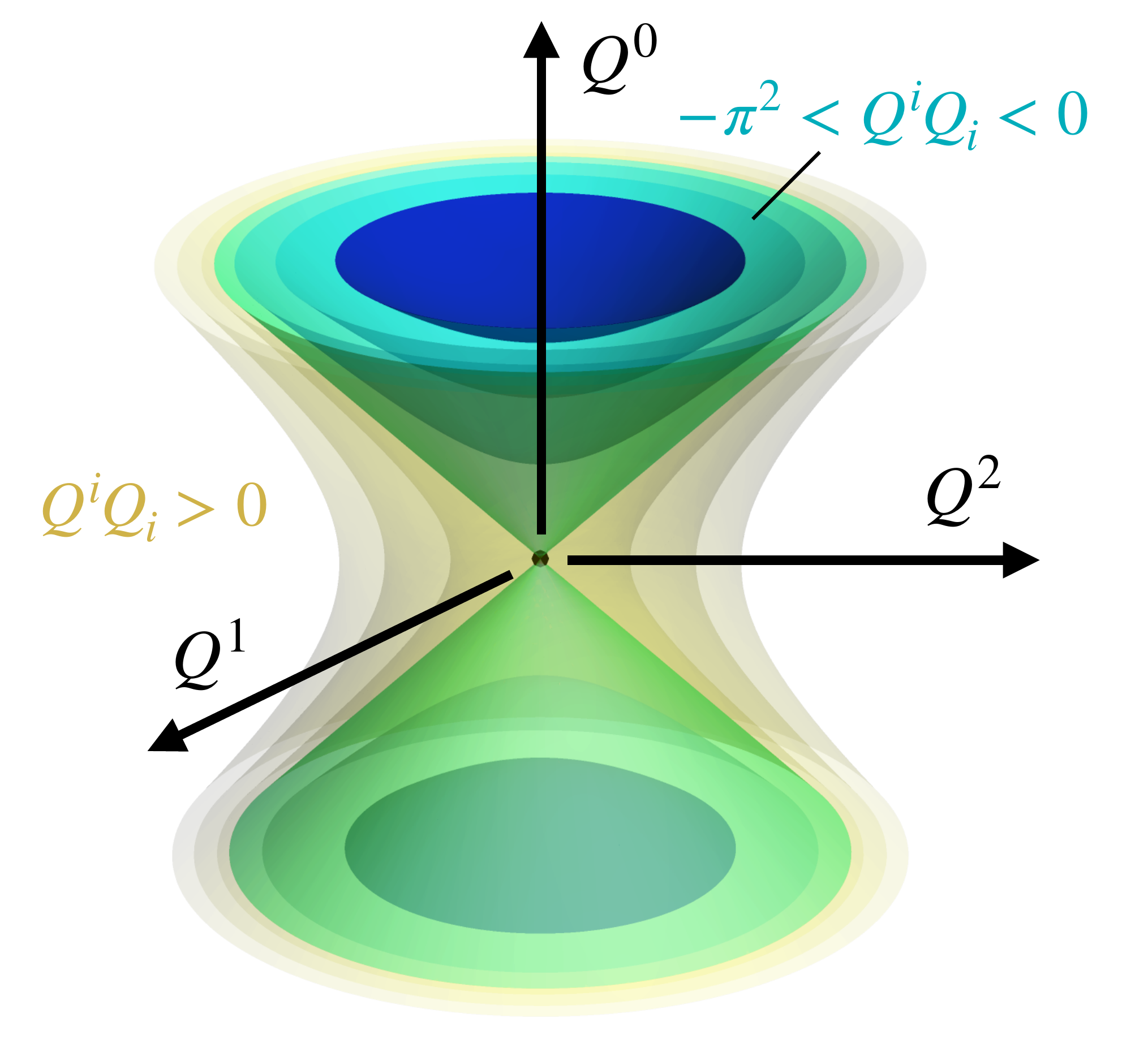}
\caption{The region $\mathcal{R}$ (\ref{R}) in three-dimensional Minkowski space is in one-to-one correspondence with elements of $\G$. The blue points correspond to $Q$ which generate rotations, the yellow points boosts, and the green points shears/null rotations.   $\mathcal{R}$ lies between the sheets of the hyperboloid $Q^iQ_i=-\pi^2$ (dark blue), with the upper sheet included but not the lower one. Figure borrowed from \cite{Alonso-Monsalve:2024oii}.}
\label{fig:psl}
\end{figure}

\begin{figure}
\centering
\includegraphics[width=0.63\linewidth]{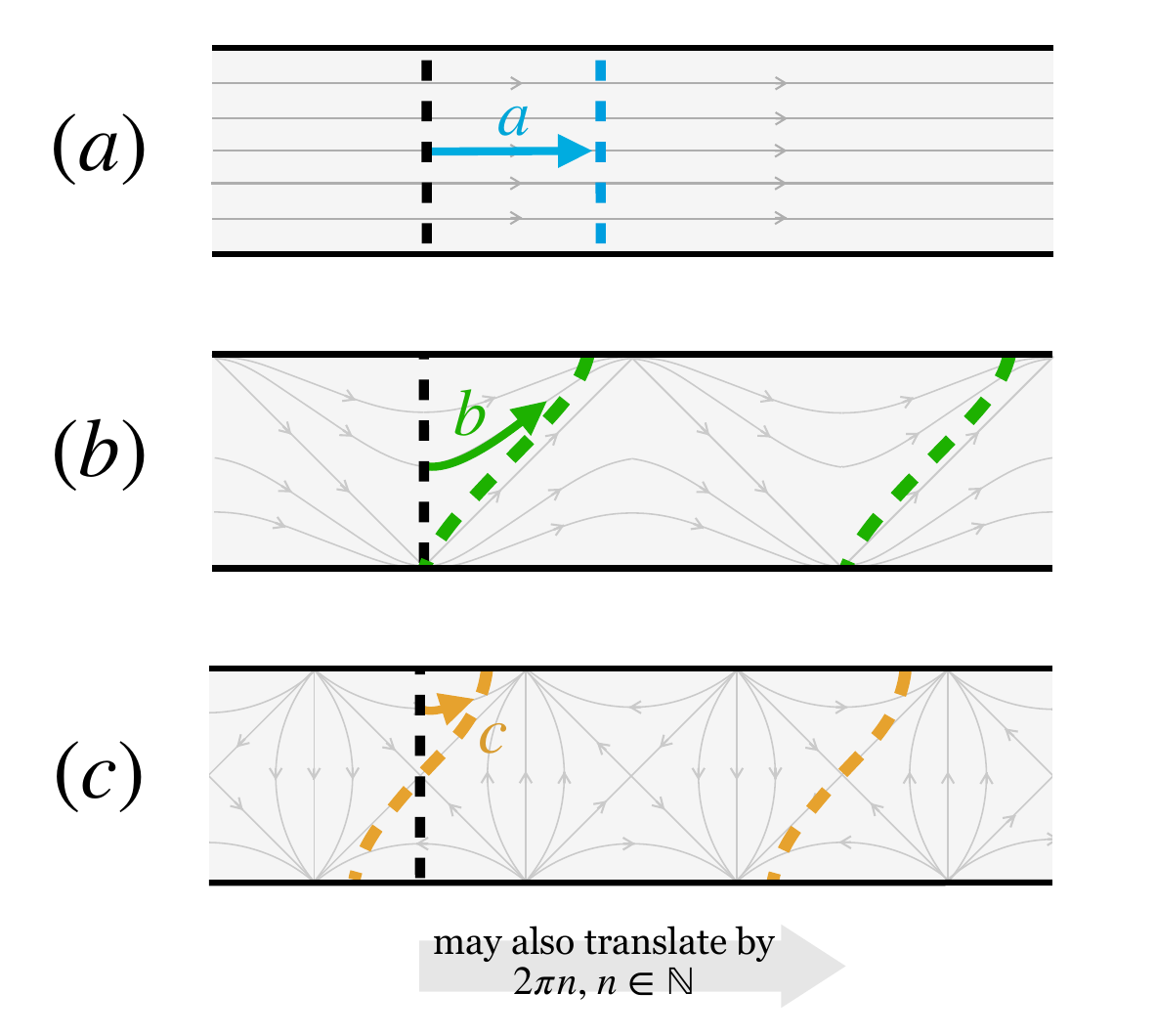}
\caption{Representative actions on the infinite strip by elements of the various orbit types of $\Gtilde$.  The colorful dashed line is the image of the black dashed line $(\sigma=0)$ under each isometry. The Killing vector fields for each isometry are shown as gray arrows. The vectors of Lie-algebra charges that generate each transformation (aside from $2\pi n$ translations) are (a) $Q^i=(a,0,0)$, timelike; (b) $Q^i=(b,-b,0)$, null; and (c) $Q^i=(0,-c,0)$, spacelike.}
\label{fig:quotients}
\end{figure}

This isometry group unfortunately does not have a matrix representation, but there exists a convenient workaround: any isometry of $q\in\Gtilde$ can be expressed in terms of an Lorentz transformation $q_0\in \G$ and an integer $n$ labeling the number of times $q$ wraps around the compact direction in $\G$. In other words, an isometry on the infinite strip corresponds to a Lorentz transformation and a $2\pi n$ translation along the $\sigma$-direction. Moreover, the exponential map from the Lie algebra $\mathfrak{g}=\mathfrak{sl}(2,\R)$ to $\G$ is surjective, so for every $q_0\in \G$ there exists a vector $Q^i$ of Lie-algebra charges such that
\begin{equation}\label{Ti}
    q_0 = e^{Q^iT_i},
\end{equation}
where $T_i$ are the generators, which we take explicitly to be
\begin{equation}   T_0=\begin{pmatrix}0&0&0\\0&0&-1\\0&1&0\end{pmatrix} \qquad T_1=\begin{pmatrix}0&0&-1\\0&0&0\\-1&0&0\end{pmatrix} \qquad
T_2=\begin{pmatrix}0&1&0\\1&0&0\\0&0&0\end{pmatrix},
\end{equation}
for concreteness. The Lie algebra endows the space of vectors $Q^i$ with a metric (given by the Killing form, up to an arbitrary normalization),
\begin{equation}
    Q_a\cdot Q_b = -Q_a^0Q_b^0 + Q_a^1Q_b^1 + Q_a^2Q_b^2,
\end{equation}
so $Q^i$ are vectors in an abstract Minkowski space. By restricting the domain of the exponential map to a subset of this Minkowski space, we can make it injective, too. In particular, if we restrict the domain to a region
\begin{equation}\label{R}
    \mathcal{R} := \{Q^i\:|\:Q\cdot Q\geq -\pi^2, Q^0>0 \:\,\mathrm{if}\:\, Q\cdot Q=-\pi^2\}\subset \R^{(1,2)},
\end{equation}
each $Q^i$ maps uniquely to a single $q_0\in G$, while the map is still surjective. See Figure \ref{fig:psl} for a depiction of this region. This gives a bijection
\begin{equation}\label{qcong}
    q \leftrightarrow (Q^i,n)
\end{equation}
so we can label every isometry $q\in\Gtilde$ by a Minkowski vector $Q^i\in\mc{R}$ and an integer $n\in\N$. See Figure \ref{fig:quotients} for an illustration of the possible quotients.

\begin{figure}
\centering
    \includegraphics[width=0.63\linewidth]{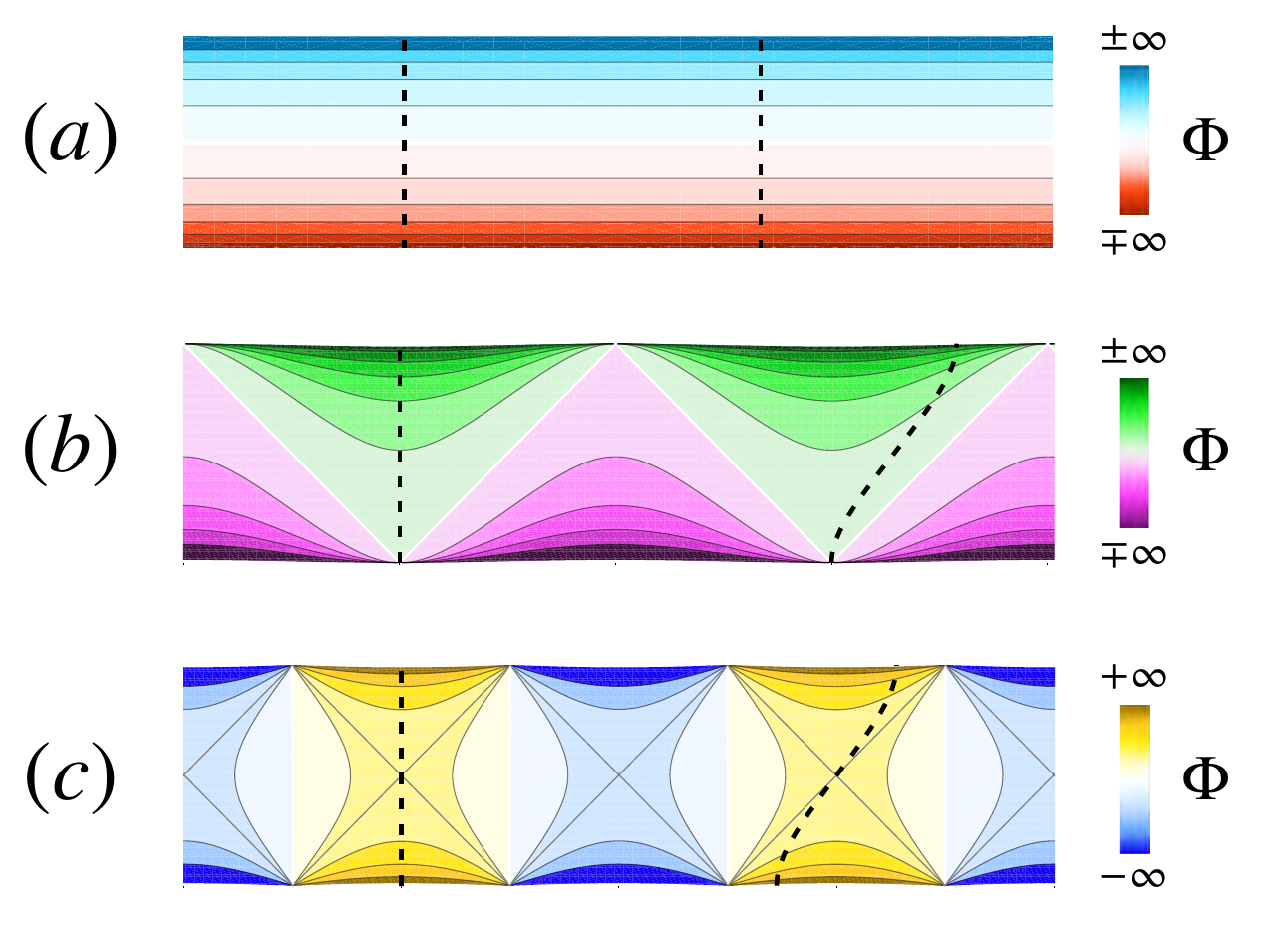}
\caption{Dilaton solutions on the geometries from Figure \ref{fig:quotients}. The dashed lines are identified under each quotient isometry. We interpret the dilaton as a proxy for volume in 2 spacetime dimensions, as suggested by its connection with near-extremal black hole limits in higher dimensions (see a discussion for e.g.~in \cite{Alonso-Monsalve:2024oii}). The regions where $\Phi\rw +\infty$ are expanding dS regions, while the regions where $\Phi\rw -\infty$ are curvature singularities. Then (a) and (b) represent expanding or crunching solutions, while (c) represents $n$ black holes which alternate with inflating regions.}
\label{fig:dil}
\end{figure}

The most general solution to the dilaton equation of motion (\ref{eoms}) is
\begin{equation}\label{Phisol}
    \Phi = V_0 \tan\tau + V_1\frac{\cos\sigma}{\cos\tau} + V_2\frac{\sin\sigma}{\cos\tau}
\end{equation}
for three parameters $V_i$. In order for the dilaton to wrap smoothly around the spacetime generated by a quotient of the infinite strip by $q$, (\ref{Phisol}) must respect this quotient isometry. In other words, we need
\begin{equation}
    \Phi(q\cdot x) = \Phi(x)
\end{equation}
for every spacetime point $x$. This translates to the requirement that
\begin{equation}
    Q^i\propto V^i.
\end{equation}
We plot the dilaton solutions in Figure \ref{fig:dil}. The vectors $Q^i$ label points in the Lie algebra $\mathfrak{g}$, which is the tangent space $T_q \Gtilde$ at each $q$.
Thus, the dilaton solutions make up the cotangent space $T^\ast_q \Gtilde$ of each isometry $q\in\Gtilde$, and therefore the unreduced phase space (also known as pre-phase space) is a cotangent bundle $T^\ast \Gtilde$. We will not discuss the symplectic structure here, and instead refer the interested reader to \cite{Alonso-Monsalve:2024oii}.

\subsection{Gauge group, orbits and stabilizers}
The cotangent bundle $T^\ast\Gtilde$ is the \textit{unreduced} phase space because we have not yet accounted for the gauge symmetries. In particular, gauge transformations act on $\Gtilde$ by the extended adjoint action $\G\rtimes\Z_2$ via
\begin{equation}\label{GZ2}
    (h,\e)\cdot q = h q^\e h^{-1}
\end{equation}
for $h\in \G$ and $\e=\pm 1$. Note that $2\pi n$ translations commute with all $g\in\Gtilde$, so we only need to consider $h\in \G$. This group action combines conjugation with inversions. This action corresponds to gauge transformations because $q$ and $(h,\e)\cdot q$ label the same quotient of the infinite strip, up to diffeomorphism. This is discussed in detail in \cite{Alonso-Monsalve:2024oii}. The upshot is that the only gauge-invariant information carried by the vector $Q^i$ is its magnitude $\sqrt{|Q\cdot Q|}$, along with its signature: timelike $Q\cdot Q < 0$, null $Q\cdot Q=0$, or spacelike $Q\cdot Q>0$. These are the only properties which remain invariant under gauge transformations, which act on $Q^i$ by rotating, boosting it and flipping it about the origin. The elements $q\in\Gtilde$ for which $Q$ is timelike are usually called elliptic; for null $Q$, parabolic; for spacelike $Q$, hyperbolic; and for zero $Q$, we will call these identity elements. These are the four orbit type strata. In Figure \ref{fig:quotients} we illustrate  these quotients. We now explain this in more detail.

The Lie group $\Gtilde$ is semisimple and thus unimodular, and therefore the Haar measure $\D q$ is invariant under conjugation and inversions, or, equivalently, under the action of the gauge group $\G\rtimes \Z_2$ (\ref{GZ2}). This means that dS-JT satisfies the assumptions from Section \ref{sec:mod}, listed above (\ref{unitarity}). Specifically, the gauge group
\begin{equation}\label{GJT}
    G=\G\rtimes \Z_2
\end{equation}
is Lie and unimodular (since $\G$ is unimodular), the unreduced phase space is a cotangent bundle $T^\ast X$ with
\begin{equation}\label{XJT}
    X = \Gtilde,
\end{equation}
and this space has a $G$-invariant measure. We can therefore apply our renormalized group averaging quantization. Group averaging requires renormalization in this theory because the standard group-averaging integral suffers the divergence described in Section \ref{sec:mod}. In particular, we have four orbit type strata
\begin{equation}\label{YkJT}
\begin{split}
    Y_0 &= \{\text{elliptic elements}\},\\
    Y_1 &= \{\text{hyperbolic elements}\},\\
    Z &= \{\text{parabolic elements}\},\\
    Z' &= \{\text{identity elements}\},
\end{split}
\end{equation}
with $X=Y_0\sqcup Y_1\sqcup Z\sqcup Z'$. These are, respectively, the region inside, outside, and on the lightcone, as well as the origin, in Figure \ref{fig:psl}, for all $n$. The orbit type strata $Z,Z'$ are measure-zero in $X$ and will not survive quantization; $X=Y_0\sqcup Y_1$ almost everywhere, as in (\ref{Xpart}). Thus, we get a split of (\ref{XJT}) as in (\ref{Hkinsplit}),
\begin{equation}
    L^2(X) = L^2(Y_0)\oplus L^2(Y_1).
\end{equation}
We now explain these orbit types in more detail. To parameterize $\G$, we identify points $Q^i\in\mc{R}$ in the ``past'' hyperboloid $Q\cdot Q =-\pi^2$ with $Q^0<0$ with points in the ``future'' hyperboloid $Q\cdot Q =-\pi^2$ with $Q^0>0$. This gives the quotiented $\mc{R}$ the topology of $\G$, $S^1\times \R^2$. A gauge transformation (\ref{GZ2}) acts on $q\cong (Q^i,n)$ as
\begin{equation}
    (Q^i,n) \mapsto (\e(hQ)^i,\e n),
\end{equation}
where $h$ acts on the vector $Q^i$ as a Lorentz transformation (rotations/boosts). To simplify notation, we let
\begin{equation}
    \Q := Q^iT_i
\end{equation}
for the Lie-algebra generators $T_i$ in (\ref{Ti}). For all $(Q^i,n)$ in $Y_k$ or $Z$, the stabilizer subgroup is
\begin{equation}\label{stabell}
    G_{(Q,n)} = \G_Q\times \{+ 1\},
\end{equation}
and, for $Z'$, $G_{(0,n)} = G$. Here,
\begin{equation}\label{GQ}
    \G_Q:=\{e^{\alpha \Q}\} \subset \G,
\end{equation}
for all $\alpha\in \R$, with the aforementioned identification. For example, for $Q^i = (1,0,0)$ timelike, any rotation generated by $\Q$ leaves $Q^i$ invariant, since
\begin{equation}
    e^{\alpha \Q} = e^{\alpha T_0} = \begin{pmatrix}
        1&0&0\\ 0&\cos\alpha&-\sin\alpha\\ 0&\sin\alpha&\cos\alpha
    \end{pmatrix}.
\end{equation}
For timelike $Q^i$, the identification of the ``future'' and ``past'' hyperboloids $Q\cdot Q=-\pi^2$ (the dark blue caps in Figure \ref{fig:psl}) makes the stabilizers compact. Moreover, any two vectors $Q^i,Q'^i$ with the same signature (timelike, spacelike, null, or even zero) are related by $Q^i=\bar\alpha (h Q')^i$ for some $\bar{\alpha}$ and $h$, so
\begin{equation}
    G_{(Q,n)} = \{(e^{\alpha \Q},1)\} = \{(e^{\alpha\bar\alpha (hQ')^iT_i},1)\} = \{(e^{\alpha (hQ')^iT_i},1)\} = \{(he^{\alpha \Q'}h^{-1},1)\}
\end{equation}
and thus their stabilizers are related by the group action,
\begin{equation}
    (h,1)\cdot G_{(Q,n)} = G_{(Q',n)}
\end{equation}
given by conjugation by $h$. This cannot relate the stabilizers of different orbit types, however, since Lorentz transformations cannot change the signature or norm of $Q^i$. Thus, elliptic, hyperbolic, parabolic  and identity ($Q^i$ timelike, spacelike, null, or zero) elements in $X$ (\ref{XJT}) form the four orbit type strata (\ref{YkJT}), as their stabilizers are related by conjugation.

Since elliptic elements have compact stabilizers, we denote their orbit type stratum by $Y_0$ in (\ref{YkJT}), following the notation from Section \ref{sec:mod}. The stabilizers of hyperbolic elements (in $Y_1$), however, are non-compact (boosts). These make the standard group-averaging integral diverge, so we need our renormalized approach from Section \ref{sec:mod}.

Finally, note that $Y_1$ is a disconnected space, with each connected component labeled by $n$. Therefore, it will further split into more superselection sectors upon quantization.

\subsection{Physical Hilbert space}\label{sec:HphysJT}
We can now choose test subspaces $\Phi_k\subset L^2(Y_k)$ for each orbit type stratum $Y_k$ with positive measure in (\ref{YkJT}), following the requirements of Section \ref{sec:Phi}, and define renormalized rigging maps $\etamodk$ by (\ref{etamod}). The resulting Hilbert spaces for spacetimes resulting from elliptic and hyperbolic quotients are $\Hphys^k$ (\ref{Hphysk}) for $k=0,1$, respectively. For simplicity, we will refer to these as the elliptic and hyperbolic Hilbert spaces.

As shown in Section \ref{sec:mod}, $\etamodk$ maps to $G$-invariant distributions, so physical states are class functions on $G$; that is, they are constant along orbits  $[q]=(\G\rtimes\Z_2)\cdot q$ in
\begin{equation}\label{config}
    \faktor{X}{G} = \faktor{\Gtilde}{\G\rtimes \Z_2}.
\end{equation}
Explicitly, this means that the physical states are wavefunctions whose only continuous argument is the norm of $Q^i$ (and decay sufficiently fast, so that $\etamodk$ converges, as ensured by (\ref{L1G})). The renormalized rigging map is given by
\begin{equation}\label{etamodJT}
\begin{split}
    \etamodk(\psi)[\chi] &= \int_{Y_k}
\D x\,\psi^\ast(x) \int_{G/G_x}\D \dot{g}\,\chi(g\cdot x)\\
&= \int_{Y_k}
\D q\,\psi^\ast(q) \sum_{\e=\pm 1}\int_{\G/\G_Q}\D \dot{h}\,\chi(hq^\e h^{-1})
\end{split}
\end{equation}
for all $\psi,\chi\in\Phi_k$, for each $k$, with $\G_Q$ defined in (\ref{GQ}). The well known Haar measures on $\G$ allow for explicit computations of the rigging map and its associated inner product (\ref{modip}) in this theory. In Appendix \ref{app:posdef} we show that the rigging map can be simplified to (\ref{etamodsimp})
\begin{equation}
    \etamodk(\psi)[\chi] = \int_{Y_k/(\G\rtimes\Z_2)}\D[q]\left(\int_{[q]}\D h\,\psi^\ast(h)\right)\left(\int_{[h]}\D m\,\chi(m)\right)
\end{equation}
where $Y_k/(\G\rtimes\Z_2)$ is the space of orbits for $q\in Y_k$ and $\D[q]$ is the measure it inherits from $Y_k$, via the pushforward of the quotient map $Y_k\rw Y_k/(\G\rtimes\Z_2)$. The inner product induced by $\etamodk$ (\ref{modip}) is Hermitian, real, and positive-definite, as shown in Section \ref{sec:mod}.

The hyperbolic Hilbert space (\ref{Hphysmod}) further splits into a countably infinite number of superselection sectors, because $\im\etamodk$ has $|\N|$ disconnected components when $k=1$, one for each natural number, generated by test functions with support on points $q$ with the same $|n|\in\N$. These are geometries with different $n$ in Figure \ref{fig:quotients}; one can see explicitly that a geometry with $Q\cdot Q\geq 0$ and one value of $n$ cannot be continuously deformed into one with a different value of $n$.

The parabolic and identity orbit type strata $Z,Z'\subset X$ have measure zero in $X$. Thus, wavefunctions in the original kinematic Hilbert space $\Hkin=L^2(X)$ which differ in their values only on $Z\cup Z'$ are considered equivalent in $\Hkin$. Therefore, $\Hkin$ cannot see the parabolic or identity orbits. Then, the physical Hilbert space of the quantum theory is given by
\begin{equation}\label{HphysJTn}
    \Hphys = \Hphys^0\oplus \left( \bigoplus_{n=0}^\infty\Hphys^{1,n} \right),
\end{equation}
where $n$ labels the superselection sectors within the hyperbolic Hilbert space $\Hphys^1$. From Figure \ref{fig:dil} we see that the elliptic Hilbert space $\Hphys^0$ contains superpositions of states which expand or crunch (as indicated by the dilaton), and the hyperbolic Hilbert space sectors $\Hphys^{1,n}$ with $n\geq 1$ contain superpositions of states with $n$ black holes. The hyperbolic Hilbert space sector $\Hphys^{1,0}$ contains superpositions of states with a conical singularity in the future or past, resulting from a quotient of the Milne patch, as described in detail in \cite{Alonso-Monsalve:2024oii}. Thus, we write, more intuitively,
\begin{equation}
\begin{split}
    \Hphys^0 &:= \ham_{\substack{\text{expand/}\\ \text{crunch}}}\\
    \Hphys^{1,0}&:=\ham_\text{sing.}\\
    \Hphys^{1,n}&:=\ham^n_\text{BH},
\end{split}
\end{equation}
and therefore (\ref{HphysJTn}) becomes
\begin{equation}\label{HphysJT}
    \Hphys = \ham_{\substack{\text{expand/}\\ \text{crunch}}} \oplus \ham_\text{sing.} \oplus \left( \bigoplus_{n=1}^\infty \ham^n_\text{BH} \right).
\end{equation}
This is first quantization procedure which quantizes all the classical solutions of dS-JT under a unified framework. Other approaches rely on gauge-fixing conditions which cannot capture all the geometries in Figure \ref{fig:quotients}; for example \cite{Held:2024rmg} use group averaging but restrict to Cauchy slices where the scale factor and the extrinsic curvature are constant, and thus miss the black hole sectors $\ham^n_\text{BH}$ in (\ref{HphysJT}), because no such slices exist in the geometries arising from hyperbolic quotients ($Q^i$ spacelike, as in Figure \ref{fig:quotients}) with $n\geq 1$. The configuration space (\ref{config}) is topologically a comb,\footnote{Not a fishbone, as suggested in \cite{Held:2024rmg}.} as seen in Figures 7 and 8 in \cite{Alonso-Monsalve:2024oii}, where the elliptic elements make up the backbone, hyperbolic ones the teeth, and parabolic (and identity ones) connect them. Finally, we note that our renormalized group averaging can be applied to closed-universe AdS-JT as well; in that case, $X=\{\text{hyperbolic elements}\}\subset \G$ and the gauge transformations are conjugation by $\G$, so there is a single sector in $\Hphys$. Physical states can be expressed as wavefunctions depending on the norm of $Q^i$, which labels the proper length of the largest Cauchy slice.

\section{Implications for gravity}\label{sec:discussion}

Through our quantization procedure, the physical states in the Hilbert space of quantum dS-JT are images of the renormalized rigging maps (\ref{etamodJT}), and thus depend only on the norm of $Q^i$ and its signature, as explained in Section \ref{sec:HphysJT}. These are gauge-invariant quantities. However, it is often useful to write physical states as wavefunctionals of quantities which are not gauge-invariant, such as the induced metric, extrinsic curvature, or value of the dilaton on a Cauchy slice. This happens naturally in the context of Wheeler-DeWitt (WDW) quantization (see e.g.~\cite{Iliesiu:2020zld} for a partial study in closed-universe JT). It would be interesting to apply group averaging to the fully unreduced phase space (labeled by the metric and dilaton fields), in order to compare to the traditional WDW quantization explicitly. Moreover, this could help clarify how group averaging can formalize the cutting and gluing constructions (summing over intermediate states) employed in gravitational path integrals (see \cite{Witten:2022xxp,Held:2025mai} for recent reviews). We save a detailed analysis WDW quantization in JT gravity for future work.

Despite these challenges, we can still comment on one important comparison between our renormalized group averaging and WDW quantization: the inner product in the former is positive-definite---as we show in Appendix \ref{app:posdef}---, while in the latter it is not. This mismatch between quantization procedures can be puzzling, but we argue that the Klein-Gordon inner product on WDW wavefunctions only fails to be positive-definite due to a misuse of the Dirac quantization procedure: the constraint surface fails to be smooth, and this is a requirement for well-definedness of Dirac quantization.\footnote{As described in Chapter 1 of \cite{henneaux1992quantization}.} The usual fix to make this inner product positive-definite is to restrict to a positive- or negative-frequency subspace of WDW wavefunctions, and this corresponds exactly to choosing a smooth subspace of the problematic constraint surface. While we do not prove here that this inner product matches the group-averaging one, \cite{Held:2024rmg} argued that the two agree in dS-JT, at least for the first two superselection sectors in (\ref{HphysJT}).

We foresee that the divergence of the rigging map which arises from non-compact stabilizers has implications for gravity, since most classical spacetimes of interest (which arise as saddles of the path integral) have isometries, and these are exactly the stabilizer subgroups of the group of diffeomorphisms. A concrete instance is the norm of the no-boundary wavefunction for closed universes: it has been suggested that its norm in the physical Hilbert space is zero \cite{Cotler:2025gui}. Our renormalized group averaging indicates that this is an artifact from applying a rigging map with the wrong renormalization: the reason the norm of the no-boundary state seems to vanish in \cite{Cotler:2025gui} is a volume of its (non-compact) stabilizer in the denominator.\footnote{The author thanks Charlie Cummings for suggesting the connection between a divergent group-averaging rigging map and \cite{Cotler:2025gui}.} Applying our ideas of renormalized group averaging yields a finite norm. In Section \ref{sec:mod} we identified and regularized this divergence for theories where the gauge group $G$ was a finite-dimensional Lie group; this does not extend rigorously to general theories of gravity, where the group of diffeomorphisms is not even locally compact, which makes the group-averaging integral ill-defined. Nonetheless, we expect that a generalization of standard group averaging and our renormalization proposal may be possible by using path integrals.

\paragraph{Acknowledgments} I thank Chris Akers, Charlie Cummings, Lorenz Eberhardt, Guglielmo Grimaldi, Daniel Harlow, Patrick Jefferson, Dave Kaiser, Don Marolf, Sarah Racz, Manu Srivastava, and Nico Vald\'es-Meller for helpful discussions. This work was conducted in MIT's Center for Theoretical Physics -- A Leinweber Institute and supported in part by the U.~S. Department of Energy under Contract No.~DE-SC0012567.

\appendix

\section{Proof of positive-definiteness}
\label{app:posdef}

In this appendix we prove that the renormalized group-averaging inner product (\ref{modip}) is positive-definite---and thus so is the standard group-averaging inner product (\ref{ga-ip}), when it is well defined---for $G$ and $\Hkin=L^2(X)$ defined as in Section \ref{sec:mod}.

First note that we can simplify the renormalized rigging map using the orbit-stabilizer theorem. This theorem gives an isomorphism
\begin{equation}
    \alpha:\faktor{G}{G_x}\rw [x]
\end{equation}
which maps $gG_x$ to $g\cdot x$. We can use this to push forward the measure $\D x$ on $X$ to $G/G_x$. In the next few equations, we will use the standard notation $\mu_V$ to denote a measure on a space $V$ in this section---in particular $\D\mu_X(x)=\D x$ will be the $G$-invariant measure on $X$ from Section \ref{sec:mod}. The pushforward of $\alpha^{-1}$ gives a measure $\mu_{G/G_x}$ on $G/G_x$:
\begin{equation}
    \left(\alpha^{-1}_\ast\mu_X\right)(H) = \mu_X\left(\alpha(H)\right) =: \mu_{G/G_x}(H)
\end{equation}
for $H\subset G/G_x$. Notice that, for all $h\in G$ and $gG_x\in G/G_x$,
\begin{equation}
    \alpha(hgG_x) = (hg)\cdot x = h\cdot (g\cdot x) = h\alpha(gG_x),
\end{equation}
so the measure $\mu_{G/G_x}$ is $G$-invariant:
\begin{equation}
    \mu_{G/G_x}(hH) = \mu_X\left( \alpha(hH) \right) = \mu_X\left( h\alpha(H) \right) = \mu_X\left( \alpha(H) \right) = \mu_{G/G_x}(H)
\end{equation}
for all $h\in G$ and $H\in G/G_x$, by $G$-invariance of $\mu_X$. The measure $\D\dot{g}$ on $G/G_x$ in (\ref{etamod}), introduced in Section \ref{sec:mod}, is the unique $G$-invariant measure on $G/G_x$, up to constant rescalings,\footnote{See e.g.~Theorem 1.5.3 in \cite{DeitmarEchterhoffHA}.} so, up to rescaling,
\begin{equation}
    \D \dot{g} = \D\mu_X(g\cdot x),
\end{equation}
and we can rewrite the renormalized rigging map (\ref{etamod}) as
\begin{equation}
    \etamodk(\psi)[\chi] = \int_{Y_k}
\D x\,\psi^\ast(x) \int_{[x]}\D y\,\chi(y),
\end{equation}
where $\D y = \D\mu_X(y)$. We can also write the integral over $Y_k$ as
\begin{equation}
    \int_{Y_k}\D x\,F(x) = \int_{Y_k/G}\D[x]\int_{[x]}\D y\,F(y),
\end{equation}
where the quotient $Y_k/G$ is the space of orbits $[x]$ and its measure, written succinctly as $\D[x]$, is the pushforward of $\D x$ under the quotient map $Y_k\rw Y_k/G$.\footnote{This measure need not be finite (although it is $\sigma$-finite under the very mild assumption that $X$ is second-countable), but the integral over orbits $Y_k/G$ still converges for functions of $[x]$ which decay sufficiently fast, as ensured by (\ref{L1G}).} Thus, we can write $\etamodk$ as
\begin{equation}
    \etamodk(\psi)[\chi] = \int_{Y_k/G}\D[x]\int_{[x]}\D y\, \psi^\ast(y) \int_{[y]}\D z\,\chi(z).
\end{equation}
Notice that $y\in[x]$, so $[y]=[x]$, and we can separate the two orbit integrals:
\begin{equation}\label{etamodsimp}
    \etamodk(\psi)[\chi] = \int_{Y_k/G}\D[x]\left(\int_{[x]}\D y\, \psi^\ast(y)\right) \left(\int_{[x]}\D y\,\chi(y)\right).
\end{equation}
With this simpler form of the renormalized group-averaging inner product, positive-definiteness follows in a straightforward manner:
\begin{equation}
    \etamodk(\psi)[\psi] = \int_{Y_k/G}\D[x]\left|\int_{[x]}\D y\, \psi(y)\right|^2
\end{equation}
is non-negative and only vanishes when
\begin{equation}
    \int_{[x]}\D y\, \psi^\ast(y)=0,
\end{equation}
or, equivalently, when $\etamodk(\psi)$ is identically zero. This completes the proof. By (\ref{etavoletamod}), this result extends to the standard group-averaging rigging map when it is well defined.

\bibliographystyle{jhep}
\bibliography{bibliography}
\end{document}